\documentclass[conference,a4paper,table]{IEEEtran}
\IEEEoverridecommandlockouts
\usepackage[T1]{fontenc} %
\usepackage{amsmath,amssymb,amsfonts}
\usepackage{algorithmic}
\usepackage{graphicx}
\usepackage{booktabs}
\usepackage{textcomp}
\usepackage{todonotes}

\def\BibTeX{{\rm B\kern-.05em{\sc i\kern-.025em b}\kern-.08em
    T\kern-.1667em\lower.7ex\hbox{E}\kern-.125emX}}

\usepackage{array}
\newcolumntype{H}{>{\setbox0=\hbox\bgroup}c<{\egroup}@{}}

\usepackage{xcolor}
\usepackage{tabularx}
\usepackage{listings}
\definecolor{darkblue}{rgb}{0,0.1,0.4}
\definecolor{darkgreen}{rgb}{0,0.7,0.2}

\usepackage{adjustbox}

\usepackage[backend=biber,
            style=numeric,
            natbib=true,
            doi=false,
            isbn=false,
            url=true,
            maxnames=5, %
            maxcitenames=1,
            giveninits=true, %
            sorting=none, %
            bibencoding=utf8]{biblatex}

\AtEveryBibitem{\ifentrytype{online}{}{\clearfield{urldate}}} %
\usepackage{multirow}

\newcommand{\infinalpaper}[1]{{\color{red!45!black}in final paper}} %

\usepackage[pdftex,
            pdfauthor={Benjamin Herb, Steve Göring, Alexander Raake and Rakesh Rao Ramachandra Rao},
            pdftitle={How Accurate are Video Quality Models for Diffusion-Based Video Super-Resolution?}
            ]{hyperref}

\usepackage{eso-pic}
\newcommand{\conf}[1]{
\AddToShipoutPictureBG*{
\AtPageUpperLeft{
 \put(\LenToUnit{0pt},\LenToUnit{-1cm}){
     \parbox{\paperwidth}{\centering\fontsize{9}{11}\selectfont\itshape #1}}
 }}}
\newcommand{\notice}[1]{
\AddToShipoutPictureBG*{
\AtPageLowerLeft{
 \put(\LenToUnit{\oddsidemargin + 1in},\LenToUnit{1cm}){
     \parbox{\textwidth}{\centering\fontsize{7}{9}\selectfont #1}}
 }}}

\begin{document}

\conf{This work has been accepted for the 18th International Conference on Quality of Multimedia Experience (QoMEX 2026).\\The final published version will be available via IEEE Xplore.}
\notice{\copyright{} 2026 IEEE. Personal use of this material is permitted. Permission from IEEE must be obtained for all other uses, in any current or future media, including reprinting/republishing this material for advertising or promotional purposes, creating new collective works, for resale or redistribution to servers or lists, or reuse of any copyrighted component of this work in other works}

\title{How Accurate are Video Quality Models for Diffusion-Based Video Super-Resolution?}

\author{%
\IEEEauthorblockN{Benjamin Herb\IEEEauthorrefmark{1}, Steve G\"oring\IEEEauthorrefmark{1}, Alexander Raake\IEEEauthorrefmark{2}, Rakesh Rao Ramachandra Rao\IEEEauthorrefmark{1} \\
\IEEEauthorblockA{\IEEEauthorrefmark{1}Audiovisual Technology Group, Technische Universität Ilmenau, Germany}
\IEEEauthorblockA{\IEEEauthorrefmark{2}Institute for Communications Engineering, RWTH Aachen University, Germany}
Email: [benjamin.herb, steve.goering, rakesh-rao.ramachandra-rao]@tu-ilmenau.de \\ raake@ient.rwth-aachen.de
}
}

\maketitle

\begin{abstract}
Recent video super-resolution (VSR) approaches use deep neural networks to enhance low-quality input videos and recover visual detail, with diffusion-based methods in particular showing promising results.
In this paper, we investigate whether existing video quality models can be used to assess the performance of these diffusion-based VSR methods, by comparing model predictions with results from a subjective test. 
The study compares six upscaling methods (Lanczos, Rhea, SCST, DOVE, SeedVR2, Starlight Mini) applied to both compressed (AV1 and DCVC-RT) and uncompressed low-resolution videos considering the play-out on a UHD-1/4K screen.
A range of full- and no-reference quality models are used to assess their applicability to this new type of quality degradation, focusing on within-sequence performance.
The results highlight that CNN-based full-reference models, such as LPIPS, DISTS, and CVQA-FR show significantly higher correlation coefficients than both conventional full- as well as the tested no-reference models.
Most overestimate the overly sharp results of SCST, with VMAF mainly failing due to spatial inconsistencies introduced by Starlight Mini.
None of the tested video quality models reach sufficient accuracy so as to replace complementary subjective testing. 
The reference, degraded and upscaled videos, as well as the user ratings and model scores are made available with the paper\footnote{\url{https://github.com/Telecommunication-Telemedia-Assessment/AVT-VQDB-UHD-1-VSR}} as open data. \end{abstract}


\begin{figure*}
	\centering
	\includegraphics[width=\textwidth]{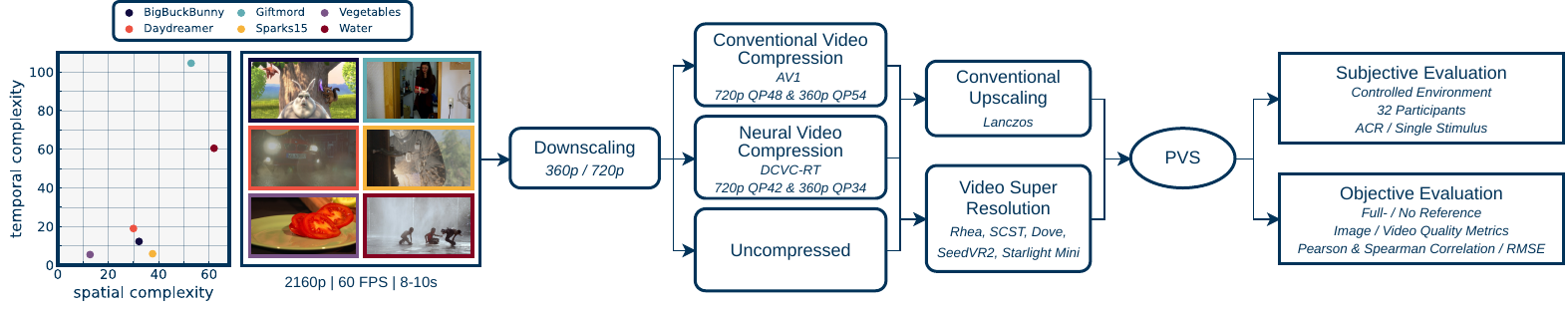}
	\caption{Overall Processing Pipeline. Spatial and temporal complexity calculated using the video complexity analyzer (VCA)~\cite{menon_vcavideocomplexity_2022}}
	\label{fig:teaser}
\end{figure*}

\section{Introduction and Related Work}
Video super-resolution methods are being developed with the goal of upscaling and enhancing low-resolution or compressed videos.
The use of deep learning for video upscaling has steadily increased over the last decade, with different architectures being employed, such as 3D CNNs, encoder-decoder structures, recurrent neural networks, and generative adversarial networks~\cite{baniya_surveydeeplearning_2024}.
More recently, diffusion models have been used, either adapting text-to-image models, such as Upscale-A-Video \cite{zhou_upscaleavideotemporalconsistentdiffusion_2024} and SCST~\cite{shi_selfsupervisedcontrolnetspatiotemporal_2025} or text-to-video models such as SeedVR \cite{wang_seedvrseedinginfinity_2025}, DOVE \cite{chen_doveefficientonestep_2025} and SeedVR2 \cite{wang_seedvr2onestepvideo_2025}.
These diffusion-based approaches are usually evaluated using conventional and learning-based quality models, with all of them utilizing PSNR, SSIM, LPIPS~\cite{zhang_unreasonableeffectivenessdeep_2018}, CLIP-IQA~\cite{wang_exploringclipassessing_2023}, and Dover~\cite{wu_exploringvideoquality_2023}. 
Most also tested with DISTS~\cite{ding_imagequalityassessment_2020}, MUSIQ~\cite{ke_musiqmultiscaleimage_2021} and NIQE~\cite{mittal_makingcompletelyblind_2013}, with DOVE additionally applying FasterVQA~\cite{wu_neighbourhoodrepresentativesampling_2023}.
To validate SeedVR2, \citeauthor{wang_seedvr2onestepvideo_2025} \cite{wang_seedvr2onestepvideo_2025} conducted a small-scale expert study, which found that the subjective results did not particularly align with the model results.

There have also been several video quality analysis studies on this topic. %
\citeauthor{he_comparativestudysuperresolution_2023} \cite{he_comparativestudysuperresolution_2023} compared five deep learning-based VSR algorithms on $4\times$ upscaling of compressed (H.264) videos.
The methods were evaluated using traditional quality models as well as a subjective study using Degradation Category Rating (DCR). %
\citeauthor{molodetskikh_aim2024challenge_2025} \cite{molodetskikh_aim2024challenge_2025} hosted the AIM Challenge on VSR Quality Assessment,
introducing a dataset generated from ten source videos that were downscaled by $2\times$ and $4\times$ and compressed using several codecs (H.264, H.265, and AV1) at various quality levels.
The resulting videos were upscaled using seven models and ranked by pairwise comparisons collected via crowdsourcing.
The submitted quality models were evaluated within-sequence and showed improvements over the baseline models PieAPP and Q-Align.
\citeauthor{borisov_vsrqadvideosuperresolution_2026} \cite{borisov_vsrqadvideosuperresolution_2026} presented a dataset which also uses $2\times$ and $4\times$ scaling combined with H.264, H.265 and AV1 compression on 20 sources.
The videos were upscaled to 1080p using eleven VSR methods and rated per source using pair comparisons and crowdsourcing.
The results for the large set of tested metrics indicated weak overall performance, with Spearman correlation coefficients below 0.68 compared to 0.84 for a compression-only set with the same sources, highlighting the different requirements for super-resolution quality assessment.

While conventional VSR methods have been evaluated in various subjective studies, none included diffusion-based methods or resolutions over 1080p.
Recent diffusion-based approaches are rapidly developing, raising the question of how to assess the quality of their outputs.
Many of these VSR methods have been evaluated only using instrumental methods during development, which might not sufficiently capture the new types of distortions, such as added details not present in the original source material. 
This paper addresses these gaps through a subjective quality evaluation using several recent VSR methods, a diverse set of source degradations, and high-resolution (4K/UHD-1) videos.
The results are used to evaluate the accuracy of existing quality models.

\section{Test Design}

To evaluate different VSR methods, we designed a subjective test.
For a realistic scenario, we apply the VSR approaches to both uncompressed and compressed source videos. 
Both a conventional and neural video codec are employed for compression to evaluate potential upscaling performance differences.
Five VSR methods are introduced, with Lanczos being included for comparison.
The suitability of quality models is assessed both for validating VSR methods on a given source sequence (within-sequence) and overall.
The overall processing pipeline is illustrated in Figure~\ref{fig:teaser}.

\subsection{Videos}

A selection of six 8-10s, 4K/UHD-1, 60\,fps source clips were used from the publicly available AVT-VQDB-UHD-1 \cite{ramachandrarao_avtvqdbuhd1largescale_2019} dataset for this test.
As a high-quality baseline, the sources were directly upscaled from 360p and 720p to 4K/UHD-1 ($3\times$ \& $6\times$) without compression artifacts to assess the performance of the models on undistorted low-resolution videos.
Additionally, to cover a range of different source distortions, two different encoding types were applied.
First, AV1 (AOMedia Project AV1 Encoder v3.12.0) encoding serves as the conventional video codec baseline, as it is widely adopted.
Second, DCVC-RT (Commit: 9b7acf7)~\cite{jia_practicalrealtimeneural_2025} is included as a recent neural video codec to evaluate whether neural-based compression influences upscaling performance. %
The constant quality parameters (see Fig. \ref{fig:teaser}) were selected for both codecs to provide two different levels of source distortions with visible coding artifacts at 360p and 720p, while maintaining comparable quality between them.

\subsection{Upscaling Methods}

\begin{figure*}
    \centering
    \includegraphics[width=\linewidth]{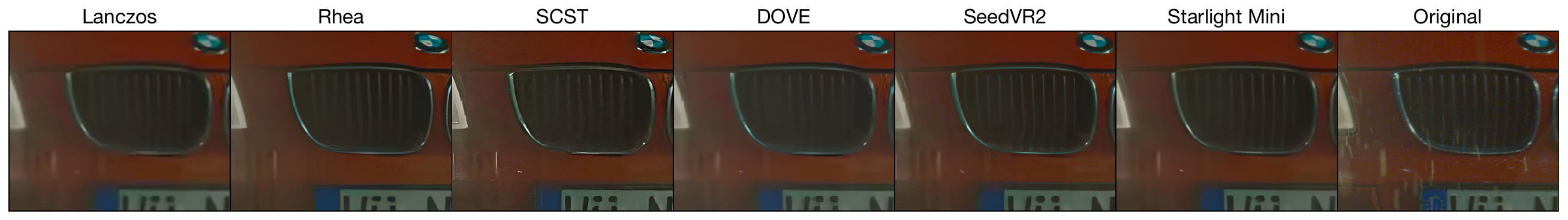}
    \caption{Example crops from the \textit{Daydreamer} sequence using 360p AV1 source encodings.}
    \label{fig:vsr_examples_long}
\end{figure*}

We selected six upscaling methods to upscale the low-resolution videos to 2160p. 
Lanczos with $a=5$ serves as the conventional upscaling reference.
For VSR, three methods (SCST, DOVE and SeedVR2) were selected from literature in addition to two commercial methods (TopazLab Rhea and Starlight Mini).
The open models were run on a 40GB A100 GPU and manually optimized to fit the memory constraints.

As first, Self-supervised ControlNet with Spatio-Temporal Continuous Mamba (SCST)~\cite{shi_selfsupervisedcontrolnetspatiotemporal_2025}, uses a text-to-image model as prior (StableDiffusion v2.1) together with spatial-temporal continuous mamba (STCM) for global 3D attention.
To leverage its text-to-image knowledge prior and align with the authors test method, Panda-70M \cite{chen_panda70mcaptioning70m_2024} is used to extract video captions for each downscaled source video. 
The model was configured using the default 20 inference steps, as well as the relatively low default temporal batch and overlap sizes of 8 and 1, as higher values lead to VRAM issues.
The variational autoencoder (VAE) is tiled using the default encoder tiling of 64 and decoder tiling of 1024 with a process size of 768.
SCST was the slowest model, with an average processing speed of 96 seconds per frame ($s/frame$).
To evaluate the visual result of SCST, Figure \ref{fig:vsr_examples_long} shows an example. 
It is visible that SCST often produces overly sharpened results. 
Additionally, the comparatively low batch size leads to noticeable temporal consistency issues.
The model's higher visible noise can mask some encoding or upscaling deficiencies. 
However, in dark areas, it occasionally produces isolated white pixels that are very noticeable.

Furthermore, we consider DOVE, proposed by \citeauthor{chen_doveefficientonestep_2025} \cite{chen_doveefficientonestep_2025},  a one-step diffusion model, which uses a text-to-video model as prior (CogVideoX).
The model is trained by first minimizing the difference between a pair of low and high-resolution images in latent space and then refining in pixel space.
During training, only the diffusion transformer is trained, while the VAE encoder / decoder weights remain frozen.
To fit VRAM, the temporal batch size is set to 128 with an overlap of 64 frames. 
For the VAE encoding / decoding, the videos are split into nine tiles with an overlap of 256 pixels.
The original code was modified to blend between spatial tiles to remove visible block boundaries, handle longer input sequences with longer temporal overlap to avoid ghosting artifacts, and prepend a number of frames to improve the quality at the start of the upscaled videos. %
This method is significantly faster (18 $s/frame$) than SCST, the outputs are smoother (see Fig. \ref{fig:vsr_examples_long}), and show strong temporal consistency due to the large batch sizes.

Furthermore, we included SeedVR2. Here, \citeauthor{wang_seedvr2onestepvideo_2025} \cite{wang_seedvr2onestepvideo_2025} use progressive distillation followed by adversarial post-training (APT) to convert a 64-step teacher diffusion model, initialized from the pretrained SeedVR diffusion transformer \cite{wang_seedvrseedinginfinity_2025}, into a one-step generator (SeedVR2).
This approach enables faster operation despite its large parameter size compared to existing multi-step models, while maintaining or improving the performance.
For this test, the largest model with 7B (16-bit) parameters is used with a temporal batch size of 25 and 12 frame overlap.
For the VAE encoding / decoding, the videos are split into nine 900x1460 tiles with 256-pixel overlap.
Smaller batch sizes again lead to significant ghosting here. 
To allow for a larger batch size, the existing code was modified to add spatial tiling to the VAE, improve temporal blending, and add a prepend frame option. %
The results (see Fig. \ref{fig:vsr_examples_long}) are slightly more detailed than DOVE and preserve smaller textures, such as the wall texture in \textit{Giftmord} better.
This was the fastest model (11 $s/frame$) among those used from literature.

Furthermore, two commercially available upscaling methods by TopazLabs\footnote{\url{https://www.topazlabs.com/}} (V7.1.0) were tested.
Rhea is one of their latest methods, which builds upon their prior Proteus and Iris models.
It provides several parameters to guide the upscaling, such as \textit{Fix compression}, \textit{Improve detail}, and \textit{Reduce noise}, which were set automatically by the tool for this test.
The model generally produced the most stable results, though it also offered less potential for detail recovery compared to the diffusion-based models.
The second Topaz model is Starlight Mini, their first diffusion-based model, which can be run locally. 
As this model only allows upscaling of $2-4\times$, the 360p source videos were upscaled first to 540p using Lanczos before scaling them to 2160p.
The tested version of this model includes spatial alignment issues, which result in parts of the image being offset slightly. 
This is not noticeable without a reference, though the performance of full-reference models might be reduced due to this.
The results are temporally stable, but generally slightly less detailed than SeedVR2.

\subsection{Quality Models}
Several full- and no-reference (FR/NR) image quality assessment (IQA) and video quality assessment (VQA) models were included in the study.
The IQA models were adapted to videos by averaging the scores sampled at two frames per second as a practical compromise between coverage and computation time.
PSNR, SSIM, and MS-SSIM typically serve as the conventional FR baseline.
Additionally, improved conventional IQA models such as PSNR-HVS, SSIMULACRA2~\cite{jonsneyers_ssimulacra2structural_2022}, and Butteraugli~\cite{jyrkialakuijala_butterauglitoolmeasuring_2016} are often used for evaluating learning-based image compression \cite{jenadeleh_subjectivevisualquality_2025}.
VQA models based on handcrafted features include VMAF (both the default and the No-Enhancement-Gain (NEG) variant) and ColorVideoVDP (CVVDP)~\cite{mantiuk_colorvideovdpvisualdifference_2024}.
Recently, CNN-based FR models such as PieAPP \cite{prashnani_pieappperceptualimageerror_2018a}, LPIPS (AlexNet and VGG)~\cite{zhang_unreasonableeffectivenessdeep_2018}, DISTS~\cite{ding_imagequalityassessment_2020}, and CompressedVQA-FR (CVQA-FR)~\cite{sun_deeplearningbased_2021} have been used more often for evaluation, with LPIPS and DISTS sometimes serving as perceptual loss functions in VSR training \cite{zhou_upscaleavideotemporalconsistentdiffusion_2024}\cite{chen_doveefficientonestep_2025}.

For NR assessment, natural scene statistic-based IQA models include BRISQUE~\cite{mittal_noreferenceimagequality_2012} and NIQE~\cite{mittal_makingcompletelyblind_2013}.
Several deep learning-based NR models are used as well, covering a range of architectures.
This includes the transformer-based IQA model MUSIQ~\cite{ke_musiqmultiscaleimage_2021} and VQA models FAST-VQA~\cite{wu_fastvqaefficientendtoend_2022} / FasterVQA~\cite{wu_neighbourhoodrepresentativesampling_2023}.
MDTVSFA~\cite{li_unifiedqualityassessment_2021} uses CNN features in combination with a recurrent neural network to model temporal memory effects, UVQ~\cite{wang_richfeaturesperceptual_2021} uses an ensemble of separately trained CNNs, while CompressedVQA-NR (CVQA-NR)~\cite{sun_deeplearningbased_2021} extracts statistics from CNN latents.
The IQA model CLIP-IQA+~\cite{wang_exploringclipassessing_2023} and VQA model MaxVQA~\cite{wu_explainableinthewildvideo_2023} rely on CLIP embeddings, though the latter incorporates FAST-VQA context for detail preservation.
Q-Align~\cite{wu_qalignteachinglmms_2024} employs an LLM for its prediction.
Dover~\cite{wu_exploringvideoquality_2023} combines a transformer for technical with a CNN for aesthetic assessment, while COVER~\cite{he_covercomprehensivevideo_2024} extends this by adding CLIP embeddings.

\subsection{Experimental Procedure}
The study was conducted with 32 participants in a controlled environment.
The 5-point absolute category rating (ACR) \cite{itu-t_p910subjectivevideo_2023} method was used with testing lasting between 45 and 60 minutes per participant, with a short break during the test.
AvrateNG\footnote{\url{https://github.com/Telecommunication-Telemedia-Assessment/avrateNG}}~\cite{goring_avratevoyageropen_2021} was used to collect the ratings and the videos were shown on an Asus XG43UQ UHD Monitor (43\,'') with a fixed viewing distance of 1.5H.
Before testing, each participant completed a FrACT10 vision test\footnote{\url{https://michaelbach.de/fract/}}.
The participants, which included students and employees of the university aged 23 to 36, were compensated for their participation.
Each participant rated all 222 PVS, presented in a random order.
To ensure the reliability of the participants, the outlier detection recommended in ITU-T P.910 \cite{itu-t_p910subjectivevideo_2023} was applied.
The Pearson correlation coefficient was calculated for each subject and the MOS, discarding participants with a $PLCC < 0.70$ and recalculating the MOS after each removal.
This threshold is slightly lower than the recommended threshold in P.910 (0.75) to account for a larger expected rating variance. 
The ratings of the 28 participants who passed the outlier detection were used for the subsequent analysis.

\section{Subjective Quality Assessment}
\begin{figure}[]
  \centering
  \begin{minipage}[b]{0.49\linewidth}
    \centering
    \includegraphics[height=0.86\linewidth]{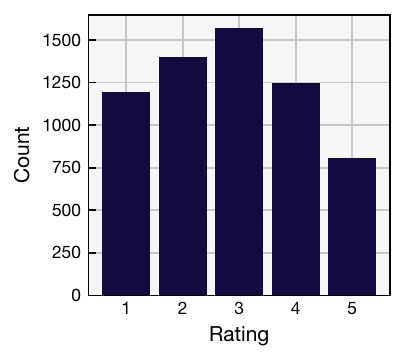}
    \caption[VSR Distribution of Ratings]{Rating Distribution}
    \label{fig:vsr_rating_distribution}
  \end{minipage}
  \begin{minipage}[b]{0.49\linewidth}
    \centering
    \includegraphics[height=0.86\linewidth]{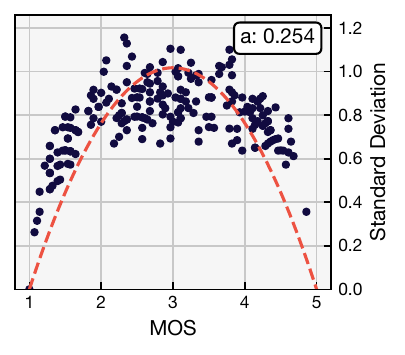}
    \caption[VSR Standard Deviation of Opinion Scores (SOS)]{SOS \cite{hossfeld_sosmosnot_2011} Analysis.}
    \label{fig:vsr_sos}
  \end{minipage}
\end{figure}

\begin{figure}
    \centering
    \includegraphics[width=\linewidth]{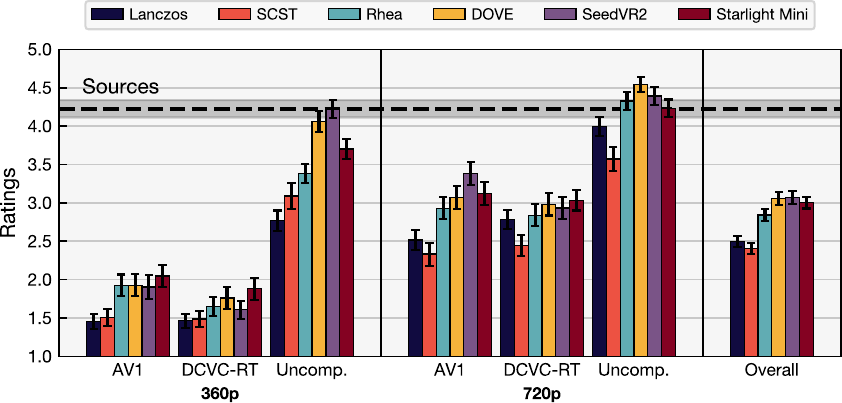}
    \caption{Overall results and results per codec / setting with 95\% CI Intervals}
    \label{fig:vsr_method_comparison}
\end{figure}

\begin{figure}
    \centering
    \includegraphics[width=1\linewidth]{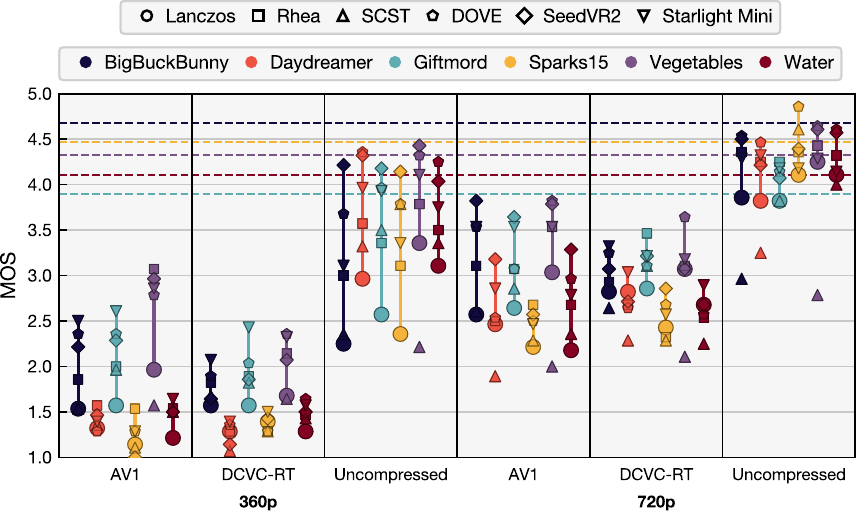}
    \caption[VSR Upsampling Method Degraded Video Comparison]{Upsampling methods results per degraded video, with the best improvements over the baseline (Lanczos) highlighted. The dashed lines show the MOS for the original source files.}
    \label{fig:vsr_method_comparison_per_source}
\end{figure}

The rating distribution in Figure \ref{fig:vsr_rating_distribution} shows an approximate normal distribution with a tendency towards lower ratings.
We furthermore conducted an SOS~\cite{hossfeld_sosmosnot_2011} analysis (see Fig. \ref{fig:vsr_sos}), and an $a$ value of 0.254 was estimated, which is within the range of similar tests, as e.g. shown in \cite{rao_largescaleevaluationsubject_2025}.

Figure \ref{fig:vsr_method_comparison} shows the results for each method, averaged over all source video sequences.
The MOS for the unaltered UHD-1 source videos is shown as a dashed line.
SeedVR2, DOVE, and Starlight Mini demonstrate the best overall upscaling performance, with none of the three significantly outperforming the others across all the tested settings.
SCST performs the worst out of the tested methods, with better performance for lower quality source videos than higher quality ones. 
This might be due to the higher amount of noise masking artifacts at the lower quality levels.
As expected, all models perform significantly better on uncompressed low-resolution videos, with SeedVR2 even achieving comparable results to the source videos when upscaling from 360p.
The rating increase from Lanczos to the three upscaled variants is noticeably higher for AV1 than for DCVC-RT. 
Figure \ref{fig:vsr_method_comparison_per_source} shows the rating changes for each of the 36 degraded videos with the improvements over the Lanczos versions being highlighted.
For the higher temporal complexity sequences (\textit{Water}, \textit{Sparks15}, \textit{Daydreamer}) at 360p, the different VSR methods only achieved minor improvements of less than 0.5 for both AV1 and DCVC-RT.
For the less temporally complex sequences (\textit{BigBuckBunny}, \textit{Giftmord}, \textit{Vegetables}), there are more considerable improvements with an increase of more than 1.0 for the AV1 encodings and a lesser increase for DCVC-RT, mirroring the overall results.
This trend continues for the compressed videos at 720p, with the upscaled AV1 sequences showing much greater improvements.
The largest improvements are for the uncompressed videos, with multiple methods matching or surpassing the perceived quality of the original UHD-1 sequences from 720p, with the results for 360p sources only being slightly lower.
It is of note here that for some sequences (e.g., \textit{Water}, \textit{Giftmord}, and \textit{Vegetable}), there is no substantial rating difference between the uncompressed 720p Lanczos scaled versions and the originals, which is likely due to the ratings being compressed from the large range of qualities in this test.
The difference in increased performance for AV1 and DCVC-RT could either be due to the methods typically being trained on conventionally compressed source material or due to different information being preserved in typical conventional codecs, though this is difficult to assess without testing more encoding types.

\section{Objective Quality Assessment}

\begin{figure}
    \centering
    \includegraphics[width=0.91\linewidth]{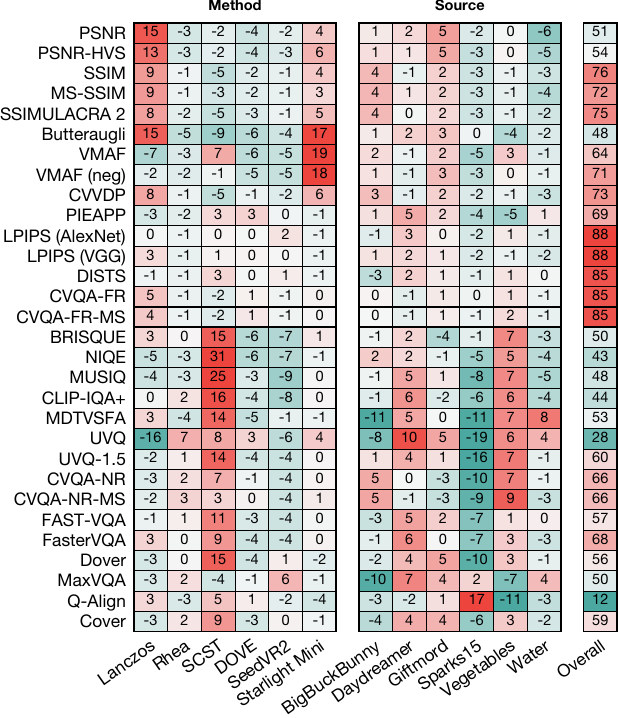}
    \caption{Spearman correlation difference ($\times 100$) without each method / source compared to the overall within-sequence result. High $\Delta$SRCC point toward a metric failing to integrate a given method / source into the overall result.}
    \label{fig:correlation_difference_matrix}
\end{figure}

\begin{figure}
    \centering
    \includegraphics[width=\linewidth]{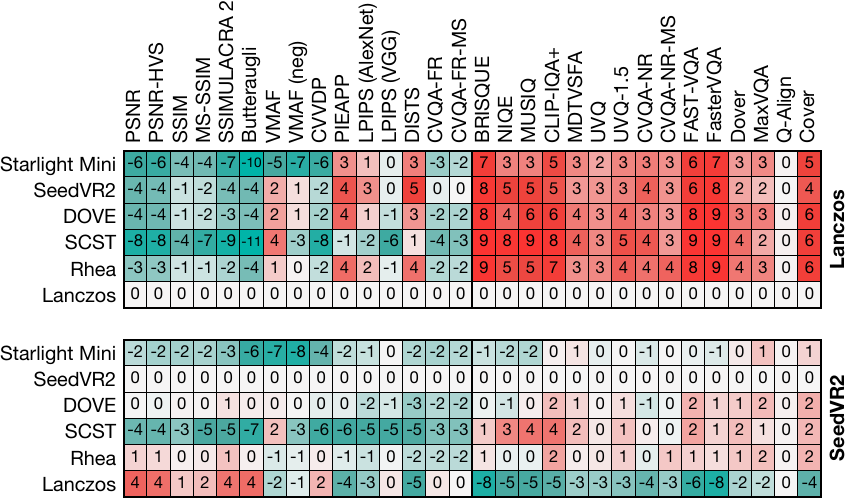}
    \caption{Prediction difference between Lanczos / SeedVR2 and each method. Red shows how much a metric overpredicts a given method [$\Delta$ MOS $\times10$].}
    \label{fig:prediction_differences_heatmap_horizontal}
\end{figure}

The resulting MOS are used to evaluate the previously introduced quality models.
Table \ref{tab:vsr_metric_correlation} shows the overall mean PLCC, SRCC, and RMSE results for the six source sequences (within-sequence) and across all PVS.
The overall result quantifies the ability of a given model to assess quality across different sequences, while the within-sequence results only focus on the quality of different versions for each source sequence separately.
Different VSR methods are typically compared applied to the same source, so for most subsequent analysis, emphasis is put on within-sequence comparisons.
For these, the resulting correlation coefficients for all six source sequences are averaged using the Fisher z-transformation to reduce sampling bias~\cite{corey_averagingcorrelationsexpected_1998}.
Furthermore, the Meng-Rosenthal-Rubin Significance Test~\cite{meng_comparingcorrelatedcorrelation_1992} is used to verify the significance of correlation coefficient differences between models, as proposed by \citeauthor{jenadeleh_subjectivevisualquality_2025}~\cite{jenadeleh_subjectivevisualquality_2025}.
Besides the overall correlation coefficients, it is important to assess how well each model integrates the different VSR methods and whether models consistently over- or underpredict them.
Figure \ref{fig:correlation_difference_matrix} shows the correlation coefficient change when removing each method / source from the set compared to the overall result. 
Furthermore, Figure \ref{fig:prediction_differences_heatmap_horizontal} visualizes the average $\Delta$MOS prediction difference to the baseline Lanczos as well as the best-performing model SeedVR2.

\begin{table}
\centering
\scriptsize
\caption[VSR Overall Results]{Mean correlation for each source (within-sequence) and overall correlation across all videos. 
The evaluation is split into FR and NR metrics, with the top three results \textbf{highlighted} and the best result \underline{underlined}.
}
\setlength{\tabcolsep}{4pt} %
\adjustbox{width=\linewidth,center}{
\begin{tabular}{l|ccc|ccc}
\toprule
 \multicolumn{1}{l}{} &\multicolumn{3}{c}{Within-Sequence}  & \multicolumn{3}{c}{Overall} \\
  Metric&PLCC & SRCC & RMSE & PLCC & SRCC & RMSE \\
\midrule
PSNR  &0.497 & 0.507 & 0.845 & 0.451 & 0.466 & 0.910 \\
PSNR-HVS  &0.510 & 0.539 & 0.841 & 0.468 & 0.476 & 0.900 \\
SSIM & 0.785 & 0.762 & 0.603 & 0.657 & 0.671 & 0.768 \\
MS-SSIM  &0.725 & 0.722 & 0.672 & 0.623 & 0.603 & 0.797 \\
SSIMULACRA 2 \cite{jonsneyers_ssimulacra2structural_2022} & 0.660 & 0.750 & 0.740 & 0.622 & 0.665 & 0.798 \\
Butteraugli \cite{jyrkialakuijala_butterauglitoolmeasuring_2016} & 0.510 & 0.481 & 0.843 & 0.509 & 0.503 & 0.877 \\
VMAF \cite{netflix_vmafvideomultimethod_2016}  &0.653 & 0.637 & 0.731 & 0.634 & 0.633 & 0.788 \\
VMAF (NEG) \cite{netflix_vmafvideomultimethod_2016} &0.715 & 0.713 & 0.680 & 0.684 & 0.684 & 0.743 \\
CVVDP \cite{mantiuk_colorvideovdpvisualdifference_2024}  &0.766 & 0.729 & 0.626 & 0.692 & 0.682 & 0.736 \\
PIEAPP \cite{prashnani_pieappperceptualimageerror_2018a} & 0.632 & 0.686 & 0.759 & 0.588 & 0.605 & 0.824 \\
LPIPS (Alex) \cite{zhang_unreasonableeffectivenessdeep_2018}  &\textbf{0.851} & \textbf{\underline{0.881}} & \textbf{0.535} & 0.609 & 0.636 & 0.808 \\
LPIPS (VGG) \cite{zhang_unreasonableeffectivenessdeep_2018}  &0.810 & \textbf{0.880} & 0.583 & 0.650 & 0.694 & 0.775 \\
DISTS \cite{ding_imagequalityassessment_2020} &0.838 &0.850 &0.551 &\textbf{0.712} &\textbf{0.712} & \textbf{0.715} \\
CVQA-FR \cite{sun_deeplearningbased_2021}  &\textbf{0.845} & 0.847 & \textbf{0.525} & \textbf{\underline{0.736}} & \textbf{\underline{0.732}} & \textbf{\underline{0.690}} \\
CVQA-FR-MS \cite{sun_deeplearningbased_2021}  &\textbf{\underline{0.856}} & \textbf{0.853} & \textbf{\underline{0.510}} & \textbf{0.733} & \textbf{0.721} & \textbf{0.693} \\
\midrule
BRISQUE \cite{mittal_noreferenceimagequality_2012}  &0.466 & 0.501 & 0.868 & 0.391 & 0.403 & 0.938 \\
NIQE \cite{mittal_makingcompletelyblind_2013}  &0.425 & 0.427 & 0.867 & 0.319 & 0.305 & 0.966 \\
MUSIQ \cite{ke_musiqmultiscaleimage_2021}  &0.500 & 0.476 & 0.813 & 0.419 & 0.440 & 0.925 \\
CLIP-IQA+ \cite{wang_exploringclipassessing_2023} &0.428 &0.441 &0.853 &0.407 &0.411 & 0.931 \\
MDTVSFA \cite{li_unifiedqualityassessment_2021} &0.451 & 0.529 & 0.826 & 0.280 & 0.263 & 0.979 \\
UVQ \cite{wang_richfeaturesperceptual_2021}  &0.327 & 0.281 & 0.886 & 0.249 & 0.215 & 0.987 \\
UVQ-1.5 \cite{wang_richfeaturesperceptual_2021}  &0.507 & 0.602 & 0.828 & 0.344 & 0.315 & 0.957 \\
CVQA-NR \cite{sun_deeplearningbased_2021}  &\textbf{0.603} & \textbf{0.665} & \textbf{0.767} & 0.450 & \textbf{0.490} & 0.910 \\
CVQA-NR-MS \cite{sun_deeplearningbased_2021}  &\textbf{\underline{0.630}} & \textbf{0.655} & \textbf{\underline{0.734}} & 0.466 & 0.488 & 0.902 \\
FAST-VQA \cite{wu_fastvqaefficientendtoend_2022}  &0.546 & 0.579 & 0.817 & \textbf{0.515} & \textbf{0.496} & \textbf{0.874} \\
FasterVQA \cite{wu_neighbourhoodrepresentativesampling_2023}  &\textbf{0.602} & \textbf{\underline{0.683}} & \textbf{0.775} & \textbf{\underline{0.556}} & \textbf{\underline{0.538}} & \textbf{\underline{0.847}} \\
Dover \cite{wu_exploringvideoquality_2023}  &0.520 & 0.549 & 0.837 & 0.388 & 0.370 & 0.939 \\
MaxVQA \cite{wu_explainableinthewildvideo_2023}  &0.430 & 0.505 & 0.868 & 0.291 & 0.314 & 0.975 \\
Q-Align \cite{wu_qalignteachinglmms_2024}  &0.116 & 0.121 & 0.949 & 0.134 & 0.108 & 1.010 \\
Cover \cite{he_covercomprehensivevideo_2024}  &0.546 & 0.592 & 0.816 & \textbf{0.480} & 0.453 & \textbf{0.894} \\
\bottomrule
\end{tabular}
}

\label{tab:vsr_metric_correlation}

\end{table}

For overall results, CVQA-FR (-MS) and DISTS significantly outperform every model besides VMAF (NEG) and CVVDP, though with a relatively low SRCC below 0.74.
None of the NR models perform well in this test, with FasterVQA achieving the highest SRCC of 0.54.

\begin{figure*}
    \centering
    \includegraphics[width=\linewidth]{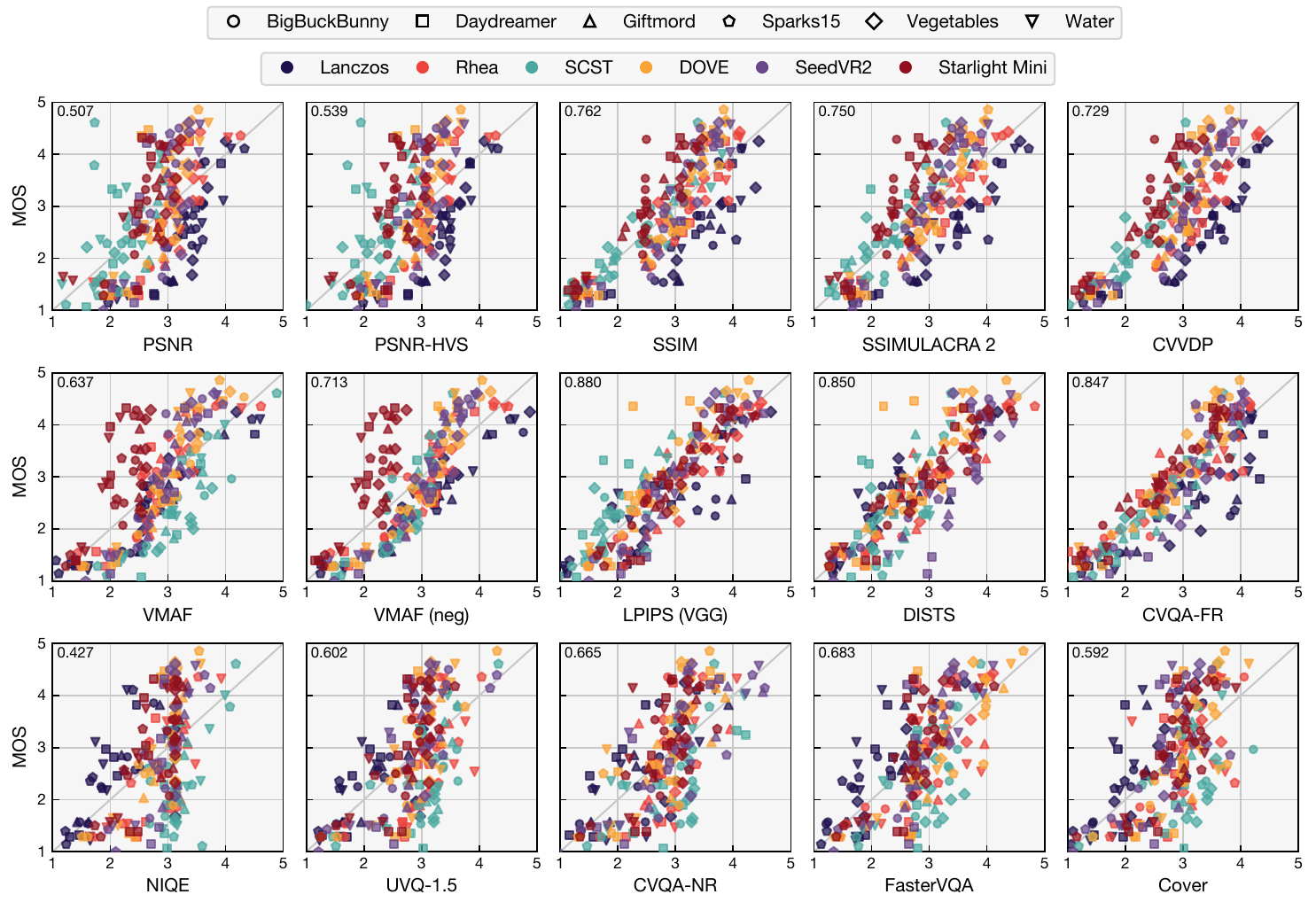}
    \caption{MOS compared to the metric results. Each metric axis is \textbf{mapped per source} to the ACR scale using third-order mapping following ITU-T Rec. P.140~\cite{itu-t_p1401methodsmetrics_2020}. The top left corner shows the mean within-sequence SRCC result.}
    \label{fig:vsr_metric_correlation_srcc_within_sequence}
\end{figure*}

For within-sequence comparisons, Figure \ref{fig:prediction_differences_heatmap_horizontal} shows a clear trend with FR models generally underpredicting the VSR results and NR models consistently overpredicting them.
LPIPS (AlexNet and VGG) shows the highest SRCC of 0.88, with CVQA-FR and DISTS achieving comparable results.
The CNN-based models (LPIPS, DISTS, CVQA-FR) significantly outperform the conventional models, likely because they are more invariant to slight texture changes introduced by the upscaling methods.
FR models that operate on the full resolution in pixel space, such as PSNR, SSIM, and especially Butteraugli and VMAF, show performance degradation due to the minor spatial inconsistencies introduced by Starlight Mini (see Fig. \ref{fig:correlation_difference_matrix}).
Also, the oversharpening of SCST gets overpredicted by VMAF, with its NEG variant successfully reducing this effect (see Fig. \ref{fig:vsr_metric_correlation_srcc_within_sequence}).
 Even though CNN-based FR models reach fairly high SRCC for within-sequence comparisons, they still show biases depending on the VSR method (see Fig. \ref{fig:prediction_differences_heatmap_horizontal}), making them unreliable for model validation.

Most NR models also struggle with the outputs of SCST, showing high SRCC improvements when removing it from the test set, with especially NIQE, MUSIQ, and CLIP-IQA+ consistently overpredicting its quality.
This highlights the importance of considering the effects of oversharpening during quality model development.
FasterVQA shows the highest mean SRCC of 0.68, with CVQA-NR (-MS) achieving similar results.
UVQ-1.5, FAST-VQA, Cover, and Dover perform slightly worse (SRCC between 0.55 and 0.60), with all of them mainly struggling to integrate the SCST results.
Neither the LLM-based VQA model Q-Align nor the CLIP-based methods that work directly with the embeddings perform well in this test, with MaxVQA performing best of the group (SRCC of 0.5).
The NR models also show much larger correlation differences depending on which source sequences are removed from the set, compared to the FR models (Fig. \ref{fig:correlation_difference_matrix}).
Removing \textit{Sparks15}, the most complex sequence, results in large performance decreases, with the opposite happening for \textit{Vegetables}, the least complex sequence.
This points towards NR models failing to differentiate between the smaller differences in low complexity scenes, which human viewers will notice, while working better on the more pronounced differences seen in high complexity scenes.
Despite NR models avoiding issues arising from details added during upscaling or other slight changes compared to the reference, none of the tested models demonstrated good enough performance to be used for VSR method validation.

\section{Conclusion}

The study presented in this paper highlights the gaps when using quality models for diffusion-based video super-resolution evaluation.
All tested models show relatively weak overall correlation.
CNN-based full-reference models outperform other architectures for within-sequence comparisons, likely because they are less susceptible to slight spatial inconsistencies introduced by some VSR methods.
However, despite avoiding these spatial alignment issues, all tested no-reference models perform significantly worse.
The results highlight the need for improved quality models, which reduce sensitivity to small spatial inconsistencies in FR models and account for oversharpening artifacts.
Due to these issues, current models are insufficient for validating new VSR methods without additional subjective testing.
Future work is needed to investigate whether the findings generalize to a broader range of source videos, encoding quality levels, and VSR methods.

\section*{Acknowledgment}
This work is part of DFG ILMETA (438822823) and ``AG Wissenschaftliches Rechnen'' of TU Ilmenau.

\section*{References}
{
\begingroup
    \renewcommand{\bibfont}{\small}
    \printbibliography[heading=none]

\endgroup
}

\end{document}